\documentclass[aip,jap,reprint]{revtex4-1}
\usepackage{bm}
\usepackage{amsmath,amssymb}
\usepackage{here}
\usepackage[dvipdf]{graphicx,color}

\begin{document}

%\title{Thermodynamic stabilization of $R$Fe$_{12}$ with $R$-site
%substitution and pressure}
\title{Effect of $R$-site substitution and the pressure on stability of $R$Fe$_{12}$: A first-principles study}

\affiliation{Research Center for Computational Design of Advanced Functional Materials, 
  National Institute of Advanced Industrial Science and Technology, 
  Tsukuba, Ibaraki 305-8568, Japan}
\affiliation{Center for Materials Research by Information Integration, MaDIS, 
  National Institute for Materials Science, Tsukuba, Ibaraki 305-0047, Japan}
\affiliation{Elements Strategy Initiative Center for Magnetic Materials, 
  National Institute for Materials Science, Tsukuba, Ibaraki 305-0047, Japan}

\author{Yosuke Harashima}
\affiliation{Research Center for Computational Design of Advanced Functional Materials, 
  National Institute of Advanced Industrial Science and Technology, 
  Tsukuba, Ibaraki 305-8568, Japan}
\affiliation{Elements Strategy Initiative Center for Magnetic Materials, 
  National Institute for Materials Science, Tsukuba, Ibaraki 305-0047, Japan}

\author{Taro Fukazawa}
\affiliation{Research Center for Computational Design of Advanced Functional Materials, 
  National Institute of Advanced Industrial Science and Technology, 
  Tsukuba, Ibaraki 305-8568, Japan}
\affiliation{Elements Strategy Initiative Center for Magnetic Materials, 
  National Institute for Materials Science, Tsukuba, Ibaraki 305-0047, Japan}

\author{Hiori Kino}
\affiliation{Center for Materials Research by Information Integration, MaDIS, 
  National Institute for Materials Science, Tsukuba, Ibaraki 305-0047, Japan}
\affiliation{Elements Strategy Initiative Center for Magnetic Materials, 
  National Institute for Materials Science, Tsukuba, Ibaraki 305-0047, Japan}

\author{Takashi Miyake}
\affiliation{Research Center for Computational Design of Advanced Functional Materials, 
  National Institute of Advanced Industrial Science and Technology, 
  Tsukuba, Ibaraki 305-8568, Japan}
\affiliation{Center for Materials Research by Information Integration, MaDIS, 
  National Institute for Materials Science, Tsukuba, Ibaraki 305-0047, Japan}
\affiliation{Elements Strategy Initiative Center for Magnetic Materials, 
  National Institute for Materials Science, Tsukuba, Ibaraki 305-0047, Japan}

%% \author{Yosuke Harashima,$^{1,3}$
%%   Taro Fukazawa,$^{1,3}$
%%   Hiori Kino,$^{2,3}$ 
%%   and 
%%   Takashi Miyake$^{1,2,3}$} 

%% \affiliation{
%%   $^1$Research Center for Computational Design of Advanced Functional Materials, 
%%   National Institute of Advanced Industrial Science and Technology, 
%%   Tsukuba, Ibaraki 305-8568, Japan\\ 
%%   $^2$Center for Materials Research by Information Integration, MaDIS, 
%%   National Institute for Materials Science, Tsukuba, Ibaraki 305-0047, Japan\\
%%   $^3$Elements Strategy Initiative Center for Magnetic Materials, 
%%   National Institute for Materials Science, Tsukuba, Ibaraki 305-0047, Japan
%% }

\date{\today}

\begin{abstract}
  We theoretically study 
  the structural stability of $R$Fe$_{12}$ %compounds 
  with the ThMn$_{12}$ structure 
  ($R$: rare-earth element, La, Pr, Nd, Sm, Gd, Dy, Ho, Er, Tm, Lu, Y, or Sc, 
  or group-IV element, Zr or Hf)
  based on density functional theory. 
  The formation energy has a strong correlation with the atomic radius of $R$. 
  The formation energy relative to simple substances decreases as the atomic radius decreases, except for $R=$ Sc and Hf,  
  while that relative to $R_{2}$Fe$_{17}$ and bcc Fe has a minimum for $R=$ Dy. 
  The present results are consistent with recent experimental reports in which the partial substitution of Zr at $R$ sites stabilizes
  $R$Fe$_{12}$-type compounds with $R=$ Nd or Sm.
  Our results also suggest that the partial substitution of Y, Dy, Ho, Er, or Tm for Nd or Sm
  is a possible way to enhance the stability of the ThMn$_{12}$ structure.
%  contribute to the stabilization of the ThMn$_{12}$ structure more than $R=$ Zr. 
  Under hydrostatic pressure, the formation enthalpy decreases up to $\approx$~6~GPa and then starts to increase 
  at higher pressures. 
%  We also show that energy gain by chemical effect is as important as 
%  that by structural change in the stabilization by Zr substitution. 
 \end{abstract}

\maketitle

\section{Introduction}
\label{Introduction}

The saturation magnetization and magnetocrystalline anisotropy are key
quantities that define the performance of a magnet compound.
A high content of iron is considered preferable for the former,
while rare-earth elements are used as a source of the latter in many cases. 
%Thus, iron-rich compounds containing small amounts of rare-earths are good candidates for strong magnet compounds.
From this viewpoint, 
$R$Fe$_{12}$-type compounds ($R$: a rare-earth element) with the ThMn$_{12}$ structure
have been studied 
as promising magnet compounds for a long time.~\cite{Bu1991a,HiMiHo2015,MiAk2018}
$R$Fe$_{12}$ contains a higher atomic percentage of Fe (92 at\%) than other magnet compounds,
e.g., Nd$_{2}$Fe$_{14}$B (82 at\%)
and $R_{2}$Fe$_{17}$ (89 at\%).
A quantitative prediction %of the large magnetization
was presented by theoretical work on NdFe$_{12}$N~\cite{MiTeHaKiIs2014}
and then confirmed by the experimental realization of NdFe$_{12}$N
by epitaxial growth.~\cite{HiTaHiHo2015}
Here, N is introduced to enhance the magnetic properties,~\cite{YaZhKoPaGe1991,YaZhGePaKoLiYaZhDiYe1991} and 
the effects of other typical elements have also been theoretically studied to improve the magnetic properties.~\cite{HaTeKiIsMi2015c,FuAkHaMi2017}

One issue is that 
$R$Fe$_{12}$ is considered to be thermodynamically unstable,~\cite{Co1990} and
partial substitution for Fe atoms is essential for
the stabilization of a bulk system with the ThMn$_{12}$ structure.
In the absence of stabilizing elements, $R_{2}$Fe$_{17}$ phases are typically generated
instead of $R$Fe$_{12}$.~\cite{MaDuDaKa1994,Su2017}
Ti is a typical stabilizing element.
SmFe$_{11}$Ti~\cite{OhTaOsSaKo1988,OhTaOsSh1988} and NdFe$_{11}$TiN~\cite{YaZhKoPaGe1991,YaZhGePaKoLiYaZhDiYe1991} were synthesized around 1990. 
A disadvantage of the introduction of stabilizing elements is the significant reduction in the saturation magnetization. 
As a matter of fact, both SmFe$_{11}$Ti and NdFe$_{11}$TiN have 
%but they are 
an inferior magnetization compared to that of Nd$_{2}$Fe$_{14}$B 
(e.g., summarized in Table~1 of Ref.~\onlinecite{HiNiMi2017}).

In order to overcome the reduction in the magnetization,
a search for another stabilizing element has been conducted.
Stabilization by substituting Fe with several elements (V, Cr, Mn, Mo, W, Al, and Si) have been reported so far,~\cite{Fe1980,YaKeJaDeYe1981,MoBu1988,OhTaOsSaKo1988,Mu1988,WaChBeEtCoCa1988}
but the reduction in the magnetization is still large.
It is theoretically suggested that 
Co can stabilize the ThMn$_{12}$ structure and retain a large magnetic moment.~\cite{HaTeKiIsMi2016}
Though a high magnetization in Sm(Fe,Co)$_{12}$ films 
has been experimentally reported recently,~\cite{HiTaHiHo2017}
thermodynamic stabilization by doping with Co has yet to be confirmed.

Recently, (Nd,Zr)(Fe,Co)$_{11.5}$Ti$_{0.5}$N$_{\alpha}$ and (Sm,Zr)(Fe,Co)$_{11.5}$Ti$_{0.5}$ were synthesized 
by the strip casting method,~\cite{SuKuUrKoSaWaKiKaMa2014,SaSuKuUrKoYaKaMa2016,KuSuUrKoSaYaKaMa2016,SuKuUrKoSaWaYaKaMa2016} 
which is more practical in industrial applications than the epitaxial growth.
Although these compounds 
contain a smaller amount of Ti 
than previously synthesized $R$Fe$_{12}$-type compounds without Zr,
the ThMn$_{12}$ structure is realized. 
%and can be generated by such synthesis process used in applications.
This suggests that the partial occupation of Zr at the $R$ sites contributes to the stabilization of the ThMn$_{12}$ structure. 
%by partially occupying the $R$ sites. %instead of Fe atoms, 
%and it is pointed out that the change in the structural stability is associated with relaxation of Fe sublattice. 
%These works suggest that 
Therefore, 
the substitution of $R$ (not Fe sites) is another possible route for stabilizing $R$Fe$_{12}$-type compounds.

In the present work, 
we theoretically examine 
a series of
$R$ elements---$R=$ La, Pr, Sm, Gd, Dy, Ho, Er, Tm, Lu, Y, Sc, Zr, and Hf---as possible stabilizing elements that
occupy the rare-earth sites in $R$Fe$_{12}$.
We calculate the formation energy of $R$Fe$_{12}$ relative to 
(i) simple substances and 
(ii) $R_{2}$Fe$_{17}$ and bcc Fe 
based on density functional theory 
and analyze the $R$ dependence.
The obtained results are discussed in connection with 
experiments on the partial substitution of Zr for $R=$ Nd or Sm. 
We then study the effect of the hydrostatic pressure in terms of the stability of $R$Fe$_{12}$. %without substitution for Fe atoms.
%based on calculation of formation enthalpy.
%Finally, the change of the formation energy induced by Zr substitution is analyzed to see if 
%the stabilization is understood from the structural change of Fe sublattices.
This paper is organized as follows. 
The computational methods are described in Sec.~\ref{Sec2}.
The calculated formation energy of $R$Fe$_{12}$ and the effect of the hydrostatic pressure are presented in 
%Calculated values of the formation energy and presenthalpy of $R$Fe$_{12}$
%are shown in
Sec.~\ref{Sec3}. 
%including the formation energy
%for formation from $R_{2}$Fe$_{17}$ and the bcc Fe.
%
%In Sec.~\ref{sec:discussion}, 
%the formation energy is divided into the contribution by structure change and that by chemical effect, 
%and analyzed in detail. 
%
%% we discuss our results in terms of partial substitution
%% for the stabilization of the NdFe$_{12}$ phase by Zr doping.
%% and validity of a previously proposed theory~\cite{SuKuUrKoSaWaYaKaMa2016} on
%% the stabilization of the $R$Fe$_{12}$ phase by Zr doping.
%for the formation energy against the $R_{2}$Fe$_{17}$ phase
%to quantify the stabilization effect of $R$-site doping with
%La, Pr, Sm, Gd, Dy, Ho, Er, Tm, Lu, Y, Sc, Zr, Hf.
%For systematic analysis to the trend of $R$-site doping, completely
%substituted systems are considered.
%Since $R$ sites are separated by a long distance, 
%the correlation among $R$ atoms is assumed to be negligible and 
%we expect that fractionally doped systems are approximately described by their constituent systems, 
%e.g. Nd$_{1-x}$Zr$_{x}$Fe$_{12}$ $\approx$ $(1-x)$NdFe$_{12}$ $+$ $x$ZrFe$_{12}$.
%In terms of the stability against the $R_{2}$Fe$_{17}$ phase,
%pressure of 6 GPa possibly stabilizes the RFe$_{12}$ phase.
%% We also discuss validity of a previously proposed theory on
%% the stabilization of the $R$Fe$_{12}$ phase by Zr doping.
%Inspired by the structural distortion due to $R$-site substitution, 
%we also investigate application of hydrostatic pressure as 
%a possible method for stabilization.
The paper is concluded in Sec.~\ref{sec:conclusion}.

\section{Calculation methods}
\label{Sec2}
We perform first-principles calculations by using QMAS (the Quantum MAterials Simulator),~\cite{Qm2014}
which is based on density functional theory~\cite{HoKo1964,KoSh1965} 
and the projector augmented-wave method.~\cite{Bl1994,KrJo1999} 
We use the Perdew--Burke--Ernzerhof (PBE) formula~\cite{PeBuEr1996} in the generalized gradient approximation (GGA)
for the exchange-correlation energy functional.
We sample 8 $\times$ 8 $\times$ 8 $k$ points, and 
the cutoff energy for the plane wave basis is set to 40.0 Ry.
The 4$f$ electrons of 
Pr, Nd, Sm, Gd, Dy, Ho, Er, and Tm atoms
are treated as spin-polarized open-core states, and
those of the Lu atom are treated as core states. 
%which would be justified because
%the hybridization of the states with other valence electrons
%is expected to be weak.
The number of occupied 4$f$ states is fixed to
2 (Pr), 3 (Nd), 5 (Sm), 7 (Gd), 9 (Dy), 10 (Ho), 11 (Er), 12 (Tm), and 14 (Lu).
The electron configuration is determined by Hund's first rule.
For light rare-earth elements (from Pr to Gd),
all 4$f$ electrons are assumed to be in minority spin states.
For heavy rare-earth elements (from Dy to Tm),
the minority spin states of the 4$f$ orbitals are fully occupied by seven electrons, and 
the other electrons are in majority spin states.
Note that the local spin moment at $R$ is antiparallel to the total spin moment.
Spin-orbit coupling is not included in the self-consistent calculation. 
The reliability of the open-core treatment has been discussed, e.g., in Ref.~\onlinecite{TaHaMiGo2018},
and we also check the reliability by calculating the formation energy with GGA $+ \; U$, as shown in Appendix~\ref{app:plusu}.

We study $R$Fe$_{12}$ with the ThMn$_{12}$ structure 
for 
$R=$ La, Pr, Nd, Sm, Gd, Dy, Ho, Er, Tm, Lu, Y, and Sc---rare-earth elements---and
$R=$ Zr and Hf---group-IV elements. 
The preferential sites for the group-IV elements are discussed
in Appendix~\ref{app:group4}.
As reference systems, $R_{2}$Fe$_{17}$ with the Th$_{2}$Zn$_{17}$ structure,  
$R_{2}$Fe$_{17}$ with the Th$_{2}$Ni$_{17}$ structure, and 
%We also calculate 
the simple substances of $R$ and Fe %for reference 
are studied. 
For Fe, the bcc structure is assumed;
for $R=$ La, Pr, Nd, and Sm, the dhcp structure is assumed; and 
for $R=$ Gd, Dy, Ho, Er, Tm, Lu, Y, Sc, Zr, and Hf, the hcp structure 
is assumed.
The structures of $R$Fe$_{12}$, $R_{2}$Fe$_{17}$, and the simple substances are computationally optimized. 
The calculation well-reproduces the experimental lattice constants for existing crystals, 
e.g., Sm$_{2}$Fe$_{17}$.~\cite{CoSu1990,KoFu2000,HiPaOhHo2016}
The calculated lattice constants and the inner coordinates of $R$Fe$_{12}$ and $R_{2}$Fe$_{17}$ are shown 
in Sec.~SA of the Supplemental Material.~\cite{supplementalmaterial}
From the obtained structures of the simple substances, 
we deduce the atomic radii ($r_{R}^{\text{calc}}$
and $r_{\text{Fe}}^{\text{calc}}$) of the elements: 
%Specifically speaking,
half of the shortest bond lengths is used 
as the atomic radii. 
These values are tabulated 
in Sec.~SB of the Supplemental Material.~\cite{supplementalmaterial}

\section{Results and Discussion}
\label{Sec3}

%% \subsection{Formation energy of $R$Fe$_{12}$ ($R=$ La, Pr, Nd, Sm, Gd, Dy, Ho, Er, Tm, Lu, Y, Sc, Zr, Hf) and lattice parameters}
%\subsection{Formation energy of $R$Fe$_{12}$ and lattice constants}
%\label{sec:rareearth_stability}
\begin{figure}[ht]
  \includegraphics[width=\hsize]{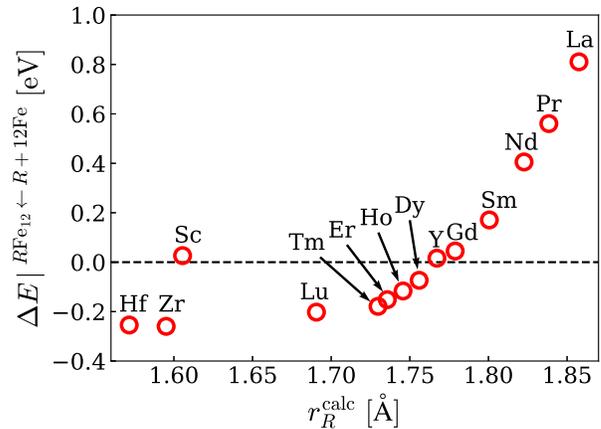}
  \caption{(Color online) Values of the formation energy $\Delta E \mid^{R\text{Fe}_{12} \leftarrow R+12\text{Fe}}$ 
    defined in Eq.~\eqref{eq:formationenergy_rfe12} 
    for $R=$ La, Pr, Nd, Sm, Gd, Dy, Ho, Er, Tm, Lu, Y, Sc, Zr, and Hf
    as a function of the atomic radius $r_{R}^{\text{calc}}$.}
  \label{fig:formationenergy_atomicradius_lanthanoid}
\end{figure}
We first discuss the values of the energy for forming $R$Fe$_{12}$
from the simple substances $R$ and Fe.
%For brevity of expression, 
Let us denote the formation energy
of a substance $X$ via the chemical reaction $X' + X'' \rightarrow X$
by $\Delta E \mid^{X \leftarrow X' + X''}$.
Figure~\ref{fig:formationenergy_atomicradius_lanthanoid} shows
the values of
$\Delta E \mid^{R\text{Fe}_{12} \leftarrow R+12\text{Fe}}$ as a function
of $r_{R}^{\text{calc}}$. 
This energy is calculated by 
\begin{equation} \label{eq:formationenergy_rfe12}
  \Delta E \mid^{R\text{Fe}_{12} \leftarrow R+12\text{Fe}}\;
  \equiv E[R\text{Fe}_{12}] - (E[R] + 12 E[\text{Fe}]),
\end{equation}
where $E[\cdot]$ denotes the total energy of the system in brackets 
per formula unit.
There is a trend toward a decrease in the formation energy as
the calculated atomic radii ($r_{R}^{\text{calc}}$)  decreases. 
When $r_{R}^{\text{calc}}$ is small, however, this trend does not hold. 
The value for Sc is exceptionally high. 
The value for Hf is slightly higher than that for Zr, although Hf has a smaller $r_{R}^{\text{calc}}$ than Zr.
Although we do not explicitly show other factors, e.g., the valency, than the atomic size, 
they may affect the results.

\begin{figure}[ht]
  \includegraphics[width=\hsize]{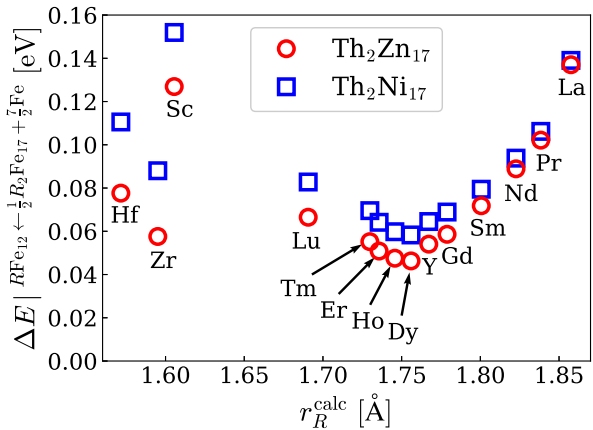}
  \caption{(Color online) Values of the formation energy $\Delta E \mid^{R\text{Fe}_{12} \leftarrow \frac{1}{2}R_{2}\text{Fe}_{17}+\frac{7}{2}\text{Fe}}$
    defined in Eq.~\eqref{eq:energydifference_1-12_2-17_1}
    for $R=$ La, Pr, Nd, Sm, Gd, Dy, Ho, Er, Tm, Lu, Y, Sc, Zr, and Hf 
    as a function of the atomic radius $r_{R}^{\text{calc}}$.
    The formation energies relative to $R_{2}$Fe$_{17}$ with the
 (rhombohedral) Th$_{2}$Zn$_{17}$ structure and
 those with the (hexagonal) Th$_{2}$Ni$_{17}$ structure
    are shown.
%    (Labels of $R$ are shown only for data points corresponding to
%    rhombohedral Th$_{2}$Zn$_{17}$-type $R_{2}$Fe$_{17}$ compounds)
 }
  \label{fig:formationenergy_atomicradius_lanthanoid_from-2-17}
\end{figure}

Considering the experimental indication that the $R_{2}$Fe$_{17}$ phase is a competing phase, 
we also calculate the formation energy of $R$Fe$_{12}$
%In order to examine effect of $R$ in $R$Fe$_{12}$ on the formation
%against the $R_{2}$Fe$_{17}$ phase,
%we consider the formation energy of $R$Fe$_{12}$ from
relative to 
$R_{2}\text{Fe}_{17}$ and bcc Fe, 
%$\frac{1}{2}R_{2}\text{Fe}_{17}+\frac{7}{2}\text{Fe}$, per $R$ atom.
%The formation energy is expressed
%as follows:
defined by 
\begin{eqnarray}
 && \Delta E \mid^{R\text{Fe}_{12} \leftarrow \frac{1}{2}R_{2}\text{Fe}_{17}+\frac{7}{2}\text{Fe}} 
  \nonumber
  \\
 && \hspace{15pt} \equiv 
  E[R\text{Fe}_{12}] - \left(\dfrac{1}{2} E[R_{2}\text{Fe}_{17}] + \dfrac{7}{2}E[\text{Fe}]\right).
  \label{eq:energydifference_1-12_2-17_1}
%  \\
% &&
%   \hspace{15pt} = 
%   \Delta E \mid^{R\text{Fe}_{12} \leftarrow R+12\text{Fe}}
%   -
%   \Delta E \mid^{\frac{1}{2}R_{2}\text{Fe}_{17} \leftarrow R+\frac{17}{2}\text{Fe}},
%  \label{eq:energydifference_1-12_2-17_2}
\end{eqnarray}
%where 
%\begin{eqnarray}
%  && \Delta E \mid^{\frac{1}{2}R_{2}\text{Fe}_{17} \leftarrow R+\frac{17}{2}\text{Fe}}
%  \nonumber
%  \\
%  && \hspace{30pt} \equiv 
%  \dfrac{1}{2}E[R_{2}\text{Fe}_{17}] - \left(E[R] + \dfrac{17}{2}E[\text{Fe}]\right).
%  \label{eq:formationenergy_2-17}
%\end{eqnarray}
%Eqs.~\eqref{eq:energydifference_1-12_2-17_1}, \eqref{eq:energydifference_1-12_2-17_2} and \eqref{eq:formationenergy_2-17} are defined for 
%
%The expression of Eq.~\eqref{eq:energydifference_1-12_2-17_2} indicates 
%the comparison between the tendency of the synthesis 
%for $R$Fe$_{12}$ and $\frac{1}{2}R_{2}$Fe$_{17}$ $+$ $\frac{7}{2}$Fe from $R$ $+$ 12 Fe.
%
We consider two cases for $R_{2}$Fe$_{17}$: %in Eq.~\eqref{eq:energydifference_1-12_2-17_1}
one with the  (rhombohedral) Th$_{2}$Zn$_{17}$ structure 
%and where it has
and the other with the (hexagonal) Th$_{2}$Ni$_{17}$ structure.
The energy difference between the two structures is discussed in 
Sec.~SC of the Supplemental Material.~\cite{supplementalmaterial}
Figure~\ref{fig:formationenergy_atomicradius_lanthanoid_from-2-17} shows
the values of
$\Delta E \mid^{R\text{Fe}_{12} \leftarrow\frac{1}{2}R_{2}\text{Fe}_{17}+\frac{7}{2}\text{Fe}}$
for the two cases.
The qualitative behavior is insensitive to the choice of structures.
As $r_{R}^{\text{calc}}$ decreases to $\sim$1.75 \AA, 
$\Delta E \mid^{R\text{Fe}_{12} \leftarrow\frac{1}{2}R_{2}\text{Fe}_{17}+\frac{7}{2}\text{Fe}}$ 
decreases. 
It has a minimum for $R=$ Dy and increases as $r_{R}^{\text{calc}}$ decreases further, 
which is in sharp contrast with the behavior of  $\Delta E \mid^{R\text{Fe}_{12} \leftarrow R+12\text{Fe}}$.
%
%%It is noteworthy that there are
%%some elements giving lower formation energy than Zr around $r_{R}^{\text{calc}} \approx 1.75$ \AA\; in the figure,
%%that is, Y, Dy, Ho, Er and Tm.
%% where $\Delta E \mid^{R\text{Fe}_{12} \leftarrow \frac{1}{2}R_{2}\text{Fe}_{17}+\frac{7}{2}\text{Fe}}$ takes the minimum at $R=$ Dy. 
%% They possibly
%% serve as a more efficient stabilizer than Zr.
The formation energy is positive even for $R=$ Dy.
%Admittedly, 
%one cannot make
%$\Delta E \mid^{R\text{Fe}_{12} \leftarrow \frac{1}{2}R_{2}\text{Fe}_{17}+\frac{7}{2}\text{Fe}}$ negative,
%thus, the $R$Fe$_{12}$ phase is more stable than $R_{2}$Fe$_{17}$,
This indicates that one cannot make
the $R$Fe$_{12}$ phase more stable than $R_{2}$Fe$_{17}$, and 
partial substitution for Fe is necessary for stabilizing the ThMn$_{12}$ structure. 
However, an appropriate choice of $R$ possibly reduces the necessary amount
of stabilizing elements that partially substitute for Fe
in synthesizing a material with the ThMn$_{12}$ structure.

As mentioned in Sec.~\ref{Introduction}, 
(Nd,Zr)(Fe,Co)$_{11.5}$Ti$_{0.5}$N$_{\alpha}$ and (Sm,Zr)(Fe,Co)$_{11.5}$Ti$_{0.5}$ 
have been synthesized.~\cite{SuKuUrKoSaWaKiKaMa2014,SaSuKuUrKoYaKaMa2016,KuSuUrKoSaYaKaMa2016,SuKuUrKoSaWaYaKaMa2016} 
These experiments suggest that the partial substitution of Zr for Nd and Sm enhances the stability of 
Nd(Fe,Co,Ti)$_{12}$N and Sm(Fe,Co,Ti)$_{12}$, respectively.
The question is if there is a better element than Zr 
that contributes to the stabilization of the ThMn$_{12}$ structure.
Figure~\ref{fig:formationenergy_atomicradius_lanthanoid_from-2-17} shows 
that the formation energy (Eq.~\eqref{eq:energydifference_1-12_2-17_1})
is lower for $R=$ Y, Dy, Ho, Er, and Tm than for Zr.
%% Among these elements, $R=$ Y has the closest atomic radius to Nd and Sm, 
%% and the lattice constants of YFe$_{12}$ are close to those of NdFe$_{12}$ and SmFe$_{12}$, as 
%% we will see below in Fig.~\ref{fig:latticeconstant_lanthanoid_1-12}. 
%% This is favorable for forming an alloy. 
%% Therefore, we can expect that partial substitution of Y for Nd or Sm enhances 
%% the stability of the ThMn$_{12}$ structure. 
We can expect that the partial substitution of these elements for Nd or Sm enhances 
the stability of the ThMn$_{12}$ structure more than Zr.

\begin{figure}[ht]
  \includegraphics[width=\hsize]{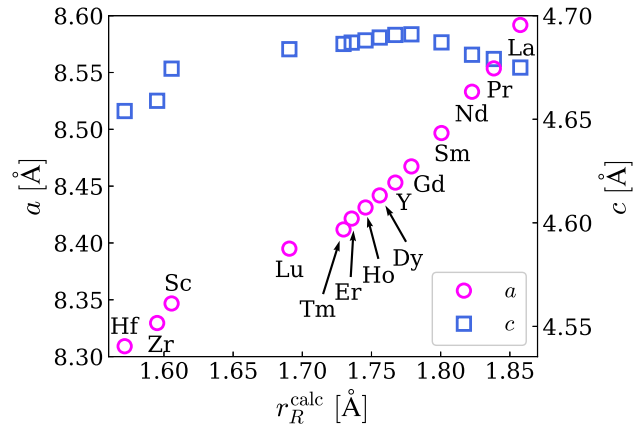}
  \caption{(Color online) Lattice constants $a$ and $c$ of $R$Fe$_{12}$ obtained by 
    structural optimization for $R=$ La, Pr, Nd, Sm, Gd, Dy, Ho, Er, Tm, Lu, Y, Sc, Zr, and Hf
    as a function of the atomic radius $r_{R}^{\text{calc}}$.
 %    (The labels of atomic symbols are for the data points of $a$).
    The left ($a$) and right ($c$) scales are taken
    so that the value corresponding to a vertical distance along the right axis becomes
    the value along the left axis multiplied by $c/a$ for $R=$ Nd.
 }
  \label{fig:latticeconstant_lanthanoid_1-12}
\end{figure}
%Here, we show the structural parameters as a function of the atomic radius.
The above results imply that the size of $R$ is essential in the stability of $R$Fe$_{12}$. 
Figure~\ref{fig:latticeconstant_lanthanoid_1-12} shows 
the lattice constants $a$ and $c$.
The $a$ axis shortens as $r_{R}^{\text{calc}}$ decreases, 
while
%the change of 
the $c$ axis is insensitive to
$r_{R}^{\text{calc}}$.
This trend is consistent with the experimental observation 
that $a$ in (Nd,Zr)(Fe,Co)$_{11.5}$Ti$_{0.5}$N$_{\alpha}$ decreases with increasing Zr concentration, 
whereas $c$ is insensitive to the Zr concentration.~\cite{SaSuKuUrKoYaKaMa2016,SuKuUrKoSaWaYaKaMa2016}
%anisotropic shrinkage observed in the experiment of partial substitution with Zr
%(The stabilization mechanism by Zr substitution is discussed in 
%Sec.~SD--SJ of the Supplemental Materials~\cite{supplementalmaterial}).
%(Change in local structure around the $R$ atom is described
%in Sec.~SC of the Supplemental Material~\cite{supplementalmaterial}).
%It is suggested in Refs.~\onlinecite{SaSuKuUrKoYaKaMa2016,SuKuUrKoSaWaYaKaMa2016} that
%the papers that 
%this shrinkage is related to the stabilization by Zr doping 
%via local distortion of the Fe sublattice around the Zr atom
%(We discuss energetics for the relaxation of the Fe sublattice in Sec.~SD--SH of the Supplemental Materials~\cite{supplementalmaterial}).
%quantification of the Fe sublattice contribution
%by decomposing the formation energy in the next section.
%However, this expectation does not agree with our result for $R=$ Sc. 
%The values of $a$ and $c$ for ZrFe$_{12}$ are not different very much from the values for $R=$ Sc, 
%whereas their formation energy are different.

%\subsection{Pressure effect}
\begin{figure}[ht]
  \includegraphics[width=\hsize]{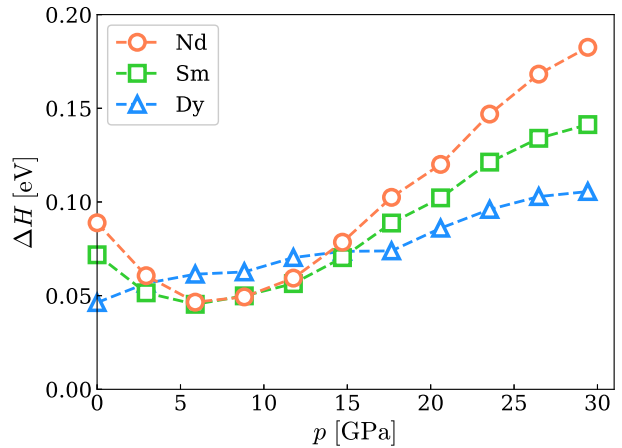}
  \caption{(Color online) Formation enthalpy defined by Eq.~\eqref{eq:formationenthalpy_pressure}
    as a function of the pressure $p$.
    The orange circles denote values for $R=$ Nd, the green squares are for $R=$ Sm, and 
    the blue triangles denote values for $R=$ Dy.}
  \label{fig:formationenthalpy_pressure}
\end{figure}
These results motivate us to consider a possibility that applied pressure stabilizes $R$Fe$_{12}$.
To discuss the stability under hydrostatic pressure, 
we estimate the formation enthalpy defined as follows:
\begin{equation}
  \Delta H(p) 
  \equiv H[R\mbox{Fe}_{12}](p) 
  - \left(
  \dfrac{1}{2}H[R_{2}\mbox{Fe}_{17}](p) +
  \dfrac{7}{2}H[\mbox{Fe}](p)
  \right),
  \label{eq:formationenthalpy_pressure}
\end{equation}
where $H[\cdot](p)$ denotes the enthalpy of the system in brackets
under pressure $p$.
At $p=$ 0, $\Delta H(0)$ is equivalent to $\Delta E \mid^{R\text{Fe}_{12} \leftarrow \frac{1}{2}R_{2}\text{Fe}_{17} + \frac{7}{2}\text{Fe}}$.
We perform computational optimization of the structure of $R$Fe$_{12}$,
$R_{2}$Fe$_{17}$ with the rhombohedral Th$_{2}$Zn$_{17}$ structure, 
and bcc Fe under hydrostatic pressure.
The applied pressure shrinks NdFe$_{12}$ anisotropically 
(see Sec. SE of the Supplemental Material~\cite{supplementalmaterial}). 
Its $a$ axis is shortened more than the $c$ axis.
The tendency of $a$ to be more susceptible than $c$ was seen also in their dependence on the atomic radius of $R$,
as seen in Fig.~\ref{fig:latticeconstant_lanthanoid_1-12}.
The $a$ axis becomes similar to that of DyFe$_{12}$ at $p \approx$~4~GPa,
and it becomes similar to that of ZrFe$_{12}$ at $p \approx$~10~GPa.
The formation enthalpy as a function of the pressure for $R=$ Nd and Sm is shown in Fig.~\ref{fig:formationenthalpy_pressure}.
As the pressure increases, the values of $\Delta H$ decrease up to $p \approx$~6~GPa 
and then start to increase when the pressure increases further.
Although applying a pressure cannot lead 
to a negative enthalpy,
it is expected that 
amount of stabilizing elements for Fe, 
such as Ti, can be reduced
by synthesis under a hydrostatic pressure of $\approx$6~GPa.
Admittedly, the pressure is too high to be applied in an industrial production process, 
the pressure is experimentally applicable and 
it may offer useful information in a viewpoint of stabilization by controlling the lattice.
We confirmed that further reduction cannot be obtained in the case of $R=$ Dy as shown in Fig.~\ref{fig:formationenthalpy_pressure}.

%% The nonmonotonic pressure dependence shown in Fig.~\ref{fig:formationenthalpy_pressure} 
The qualitative difference in the pressure dependencies shown in Fig.~\ref{fig:formationenthalpy_pressure} 
can be explained as follows.
Because we are considering a temperature of zero, the enthalpy under a finite 
pressure $H(p)$ is written as
\begin{align}
 H(p) &= H(0) + \int_{0}^{p} dp'\, V(p') \\
 &=
 H(0) + V(0)\,p - \frac{K}{2} p^2 + \cdots,
\end{align}
where $V(p)$ is the volume of the system under the pressure $p$ and
$K$ is a coefficient that is related to the bulk modulus, $B_{0}$,
by $K \equiv -d V / d p \mid_{p=0} = {V(0)}/{B_{0}}$.
%Therefore, 
In the first order of $p$, the increase in the enthalpy
is proportional to the volume at zero pressure.
It follows from the second-order term of $p$  that a phase is 
more easily stabilized by pressure when it is softer.
This can be expressed in a more general form by noting that
the higher-order terms are written as
$\int_{0}^{p} dp'\, \left( V(p') - V (0)\right)$. 
This is a general expression of  ``softness'' because
it refers to the volume change under a given pressure $p$.

At a pressure of zero, 
the gradient of $\Delta H(p)$ in Eq.~\eqref{eq:formationenthalpy_pressure}
is determined by a difference in the volume at a pressure of zero, $\Delta V_{0}$.
\begin{align}
 \Delta V_{0} &\equiv V[R\text{Fe}_{12}](0) 
 \nonumber
 \\
 & \qquad - \left(\frac{1}{2}V[R_{2}\text{Fe}_{17}](0) + \frac{7}{2}V[\text{Fe}](0)\right).
\end{align}
$V[\cdot](0)$ denotes the volume of the system in brackets per formula unit at $p=$ 0.
The values of $\Delta V_{0}$ are $-1.7$ \AA$^{3}$ for Nd, $-1.4$ \AA$^{3}$ for Sm, and $0.4$ \AA$^{3}$ for Dy, respectively.
This is consistent with the behavior of $\Delta H$ at a low pressure shown in Fig.~\ref{fig:formationenthalpy_pressure}.

In the second-order approximation,
$\Delta H(p)$ in
Eq.~\eqref{eq:formationenthalpy_pressure}
can be written as
\begin{align}
\Delta H(p) &= \Delta H_{0} + \Delta V_{0} p - \dfrac{\Delta K}{2} p^{2}
 \label{Delta_H_completing_square_prev}
\\
&=
-\frac{\Delta K}{2} \left(p - \frac{\Delta V_{0}}{\Delta K} \right)^{2}
 + \Delta H_{0}
 + \frac{\Delta V_{0}^{2}} {2\Delta K},
 \label{Delta_H_completing_square}
\end{align}
where
\begin{align}
 \Delta H_{0} &\equiv H[R\text{Fe}_{12}](0)
 \nonumber
 \\
 & \qquad - \left(\frac{1}{2}H[R_{2}\text{Fe}_{17}](0) + \frac{7}{2}H[\text{Fe}](0)\right),
 \\
 \Delta K &\equiv 
 K[R\text{Fe}_{12}]-\left(\frac{1}{2}K[R_{2}\text{Fe}_{17}]+\frac{7}{2}K[\text{Fe}]\right).
\end{align}
$K[\cdot]$ denotes the coefficient $V(0)/B_{0}$ of the system.

Within the second-order approximation,
the minimum of $\Delta H (p)$ always exists at $p>0$
when $\Delta V_{0} < 0$ and $\Delta K < 0$ hold,
or more intuitively,
when $R$Fe$_{12}$ is smaller and harder
than $R_{2}$Fe$_{17}$ at a temperature of zero.
The values estimated from our first-principles calculations are
$\Delta K = -0.070$ \AA$^{3}$/GPa
and $\Delta V_{0} = -1.7$ \AA$^{3}$ for Nd.
These values enable us to predict the existence of the dip from the information at a pressure of zero.
Note that, however,
Eq.~\eqref{Delta_H_completing_square}
well-describes the behavior of the curves in
Fig.~\ref{fig:formationenthalpy_pressure} 
only for small pressures.
The valid range does not cover 
the argument of the minimum $\Delta H(p)$,
$\Delta V_{0} / \Delta K \approx 24 \text{GPa}$.
The higher-order terms omitted in 
Eq.~\eqref{Delta_H_completing_square_prev}
take effect under high pressure 
and change the minimum of $\Delta H (p)$
from 24~GPa to 6~GPa.

%\section{Discussion}
%\label{sec:discussion}, 

\section{Conclusion}
\label{sec:conclusion}
%We have studied $R$-site substitution 
%for stabilizing the $R$Fe$_{12}$ phase 
%against the $R_{2}$Fe$_{17}$ phase.
We have performed first-principles calculations
of $R$Fe$_{12}$, % and $R_{2}$Fe$_{17}$,
where $R=$ La, Pr, Nd, Sm, Gd, Dy, Ho, Er, Tm, Lu, Y, Sc, Zr, and Hf
are considered.
The formation energy relative to simple substances 
becomes lower as the atomic radius of $R$ becomes smaller, except for $R=$ Sc and Hf. 
The stability of $R$Fe$_{12}$ relative to the $R_{2}$Fe$_{17}$ phase was also discussed. 
%in terms of formation energy.
We found that ZrFe$_{12}$ has a lower formation energy than NdFe$_{12}$ and SmFe$_{12}$.
This is consistent with the experimental results for the synthesis of (Nd,Zr)(Fe,Co)$_{11.5}$Ti$_{0.5}$N$_{\alpha}$ and (Sm,Zr)(Fe,Co)$_{11.5}$Ti$_{0.5}$.
We also found that 
%In addition, $R=$ Y, Dy, Ho, Er and Tm yield the lower formation energy 
%than that for ZrFe$_{12}$. 
%$R$Fe$_{12}$ phase
%against the $R_{2}$Fe$_{17}$ phase than Zr.
%
%Among these $R$ elements, 
Y, Dy, Ho, Er, and Tm are possible candidates for enhancing the stability 
of Nd- or Sm- based $R$Fe$_{12}$-type compounds.
%those elements can serve as efficient substitutional elements 
%in stabilizing NdFe$_{12}$ and SmFe$_{12}$.
%$R=$ Y is expected to be more efficient than Zr. Namely, smaller amount of substitution is enough to 
%obtain the same impact on the stabilization, which 
%is preferable to keep high magnetocrystalline anisotropy 
%originated from Nd- or Sm-4$f$ electrons. 
%$R=$ Dy, Ho, Er and Tm are also efficient substitutional elements in terms of stability, 
%whereas they are heavy elements, and so not favorable in terms of magnetization and industrial application. 
%Since Zr does not have 4-$f$ electrons and hardly contribute the magnetization and the magnetocrystalline anisotropy,
%Zr doping will weaken the magnetic properties in NdFe$_{12}$ and SmFe$_{12}$.
%It is expected that Y doping stabilizes with less amount than Zr 
%and performance of less doped compounds is expected to be higher.
%% Dy doping is expected to enhance the magnetocrystalline anisotropy 
%% as in Nd$_{2}$Fe$_{14}$B.
%The combination of the several elements and tuning their concentration will be important 
%for applications of permanent magnets.
The effect of hydrostatic pressure was also discussed 
in terms of the stability of the NdFe$_{12}$, SmFe$_{12}$, and DyFe$_{12}$ phases.
In the cases for NdFe$_{12}$ and SmFe$_{12}$,
a hydrostatic pressure of $\approx$6~GPa was found to contribute
to the stability of the phases, 
although the formation enthalpy is still positive.

%% We also investigated a previous theory that claims importance of
%% relaxation of the Fe sublattice caused by Zr doping. 
%% Our results suggest the importance of the remaining contributions in addition to 
%% that of the relaxation of the Fe sublattice.
%% Against the previous conjecture,
%% our results suggest that the relaxation does not contribute
%% to the stabilization
%% of the $(\text{Nd},\text{Zr})$Fe$_{12}$ phase
%% against the $(\text{Nd},\text{Zr})_{2}$Fe$_{17}$ phase.

\begin{acknowledgments}
The authors would like to thank 
Kiyoyuki Terakura, Shoji Ishibashi, 
Satoshi Hirosawa and Hisazumi Akai for fruitful discussions.
This work was supported by 
the Elements Strategy Initiative Project under the auspices of MEXT, 
by the ``Materials research by Information Integration''
Initiative (MI$^{2}$I) project of the Support Program for Starting Up Innovation Hub from
the Japan Science and Technology Agency (JST),
and also by 
MEXT as a social and scientific priority issue 
(Creation of new functional Devices and high-performance Materials to 
Support next-generation Industries; CDMSI) to be tackled by using a post-K computer.
Computations were partly carried out using the facilities of 
the Supercomputer Center, the Institute for Solid State Physics, 
the University of Tokyo, and the supercomputer of ACCMS, Kyoto University and 
the K computer provided by the RIKEN Advanced Institute for 
Computational Science (Project IDs:hp150014, hp160227, hp170100, and hp170269).
\end{acknowledgments}

\appendix
\section{Comparison between GGA $+$ open-core and GGA $+ \; U$}
\label{app:plusu}

\begin{table}[ht]
  \caption{Formation energies defined by Eqs.~\eqref{eq:formationenergy_rfe12} and \eqref{eq:energydifference_1-12_2-17_1} for 
    NdFe$_{12}$ and rhombohedral Nd$_{2}$Fe$_{17}$.
    The value of $U$ is set as 5 eV.
    The energies are presented in electronvolts.
  }
  \vspace{5pt}
  \begin{center}
    \begin{tabular}{p{105pt}rrrr}
      \hline \hline
      & \hfil Eq.~\eqref{eq:formationenergy_rfe12} \hfil & \hfil Eq.~\eqref{eq:energydifference_1-12_2-17_1} \hfil 
      \\
      \hline
      \multicolumn{1}{l}{GGA $+$ open-core} & \hfil 0.405 \hfil & \hfil 0.084 \hfil 
      \\
      \multicolumn{1}{l}{GGA $+ \; U$}         & \hfil 0.395 \hfil & \hfil 0.105 \hfil 
      \\
      \hline \hline
    \end{tabular}
  \end{center}
  \label{table:opencore-plusu}
\end{table}

In this appendix, we discuss the reliability of the open-core treatment for 4$f$ electrons 
by comparison with the calculation with the GGA $+ \; U$ method.
We calculate the total energy of NdFe$_{12}$ and rhombohedral Nd$_{2}$Fe$_{17}$ with the GGA $+ \; U$ method.
The crystal structures are optimized within the scheme.
We use 5 eV as a value of $U$ for the Nd 4$f$ orbitals.
Table~\ref{table:opencore-plusu} presents the formation energies calculated with 
Eqs.~\eqref{eq:formationenergy_rfe12} and \eqref{eq:energydifference_1-12_2-17_1}.
The formation energy calculated with the GGA $+$ open-core treatment agrees with 
the value calculated with the GGA $+ \; U$.

\section{Substitution of group-IV elements (Zr, Ti, and Hf)}
\label{app:group4}
\begin{figure}[ht]
  \includegraphics[width=\hsize]{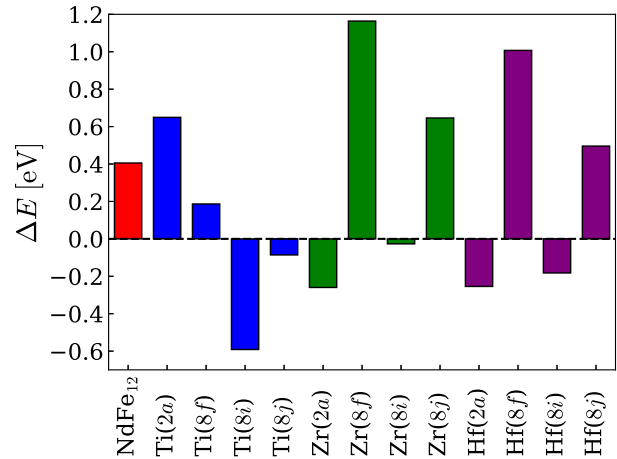}
  \caption{(Color online) Values of the formation energy for NdFe$_{12}$,
 $Z$Fe$_{12}$ ($2a$ substitution), and NdFe$_{11}Z$ (8$f$, 8$i$, and 8$j$ substitution)
 defined by Eqs.~\eqref{eq:formationenergy_zfe12} and \eqref{eq:formationenergy_ndfe11z}
 for $Z=$ Ti, Zr, and Hf.
    The labels along the horizontal axis except ``NdFe$_{12}$'' denote the substituting element
    and the site of the substitution. 
    The label Zr(2$a$), for example, corresponds to ZrFe$_{12}$ and Ti(8$i$) to
    NdFe$_{11}$Ti with one of the Fe(8$i$) sites replaced by Ti.}
  \label{fig:formationenergy_zrtihf}
\end{figure}

In order to
investigate the site preference of group-IV elements  $Z=$ Ti, Zr, and Hf for the substitution in NdFe$_{12}$,
we calculate $Z$Fe$_{12}$ ($2a$ substitution) and
NdFe$_{11}Z$ (8$f$, 8$i$, and 8$j$ substitution) with structure optimization.
Then, we calculate their formation energies from the simple Fe, Nd, and $Z$ phases.
In the case of $2a$ substitution, we consider the formation energy of 
$Z$Fe$_{12}$ plus the simple Nd phase from the simple Fe, Nd, and $Z$ phases:
\begin{equation}
 \Delta E \mid^{Z\text{Fe}_{12} + \text{Nd} \leftarrow \text{Nd} + Z + 12\text{Fe}}
 =
 \Delta E \mid^{Z\text{Fe}_{12} \leftarrow Z + 12\text{Fe}}.
 \label{eq:formationenergy_zfe12}
\end{equation}
In the other cases,
we consider the formation energy of 
NdFe$_{11}Z$ plus the simple Fe phase from the simple Fe, Nd, and $Z$ phases:
\begin{equation}
 \Delta E \mid^{\text{NdFe}_{11}Z + \text{Fe} \leftarrow \text{Nd} + Z + 12\text{Fe}}
 =
 \Delta E \mid^{\text{NdFe}_{11}Z \leftarrow \text{Nd} + Z + 11\text{Fe}}.
 \label{eq:formationenergy_ndfe11z}
\end{equation}
Therefore, we nominally use
the common reference system (Nd+$Z$+12Fe) to the cases.
%
%For the formation of (Nd$_{1-x}Z_{x}$)(Fe$_{11+x}Z_{1-x}$) by the simple substances ($Z=$ Zr, Ti, Hf), 
%the energy is defined as,
%\begin{align}
%  & \Delta E \mid^{(\text{Nd}_{1-x}Z_{x})(\text{Fe}_{11+x}Z_{1-x}) + x\text{Nd} + (1-x)\text{Fe} \leftarrow \text{Nd} + Z + 12\text{Fe}}
%  \nonumber
%  \\
%  & \quad \equiv \left\{E[(\text{Nd}_{1-x}Z_{x})(\text{Fe}_{11+x}Z_{1-x})] \right.
%  \nonumber
%  \\ 
%  & \qquad \qquad \left. + xE[\text{Nd}] + (1-x)E[\text{Fe}]\right\} 
%  \nonumber
%  \\
%  & \qquad - \left\{E[\text{Nd}]+ E[Z] + 12 E[\text{Fe}]\right\},
%  \label{eq:formationenergy_rfe12_appendix}
%\end{align}
%where Nd denotes dhcp-Nd, $Z$ denotes hcp-$Z$, Fe denotes bcc-Fe.
%$x$=1 indicates the formation of $Z$Fe$_{12}$.
%For $x$=0, NdFe$_{11}Z$ denotes one of the Fe(8$f$), Fe(8$i$), Fe(8$j$) doped systems.
%Each of the crystal structures is numerically optimized.

The results are shown in Fig.~\ref{fig:formationenergy_zrtihf}.
Ti(8$i$), Zr(2$a$), and Hf(2$a$) are the preferential sites for substitution.
As for Hf, the 8$i$ site is as stable as the 2$a$ site. 
The values of $\Delta E$ for the substituted systems are much smaller than that for NdFe$_{12}$.
Therefore, those elements can work positively for the stabilization of the ThMn$_{12}$ phase.

\begin{figure}[ht]
  \includegraphics[width=\hsize]{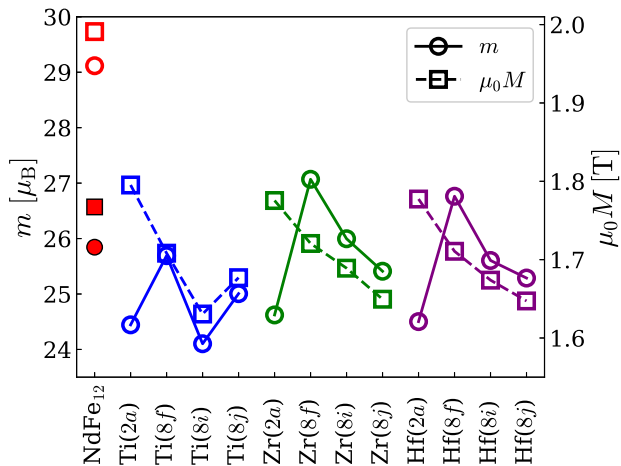}
  \caption{(Color online) Total magnetic moment [$\mu_{\text{B}}$/f.u.] (open circles) and magnetization [T] (open squares)
    for NdFe$_{12}$, $Z$Fe$_{12}$, and NdFe$_{11}Z$ ($Z=$ Ti, Zr, and Hf).
    The horizontal labels denote the same systems as in Fig.~\ref{fig:formationenergy_zrtihf}.
    The red filled circle and square denote the values for NdFe$_{12}$ without the Nd 4$f$ moment.}
  \label{fig:magneticmoment_zrtihf}
\end{figure}

We also evaluate the magnetic moments and magnetizations of the systems considered.
Figure~\ref{fig:magneticmoment_zrtihf} shows 
the magnetic moment $m$ [$\mu_{\text{B}}$/f.u.] 
and magnetization $\mu_{0}M$ [T], where $\mu_{0}$ is the vacuum permeability.
The magnetization is estimated from 
the calculated magnetic moment $m$, volume $V$, and Bohr magneton $\mu_{\text{B}}$
by $\mu_{0}M = \mu_{\text{B}} m / V$.
The values denoted by the open symbols
include 
the value of $g_{J}J=$ 3.273 $\mu_{\text{B}}$ as the contribution
from the Nd 4$f$ electrons
[$g_{J}$: Lande g-factor; $J$: total angular momentum of the Nd 4$f$
electrons].
The values denoted by the filled symbols
do not include this contribution.
%Note that the Nd-4$f$ electrons are treated as open core states in the
%present calculations.

%
%of (A) NdFe$_{12}$ with open symbols and (B) NdFe$_{11}Z$ (for $x=$ 0) include the value of $g_{J}J=$ 3.273 $\mu_{\text{B}}$ as the contribution
%from the Nd-4$f$ electrons 
%[$g_{J}$: the Lande g-factor; $J$: the total angular momentum of Nd-4$f$ electrons], 
%while values of (C) NdFe$_{12}$ with filled symbols and (D) $Z$Fe$_{12}$ (for $x=$1) do not include $g_{J}J$.~\cite{HaTeKiIsMi2015a,HaTeKiIsMi2015c}
%Note that the Nd-4$f$ electrons are treated as open core states in the present calculations. 

The magnetic moment $m$ (circle) for $Z$Fe$_{12}$ [denoted by $Z(2a)$] in the figure 
is much less than that for NdFe$_{12}$.
Though it mainly originates from the lack of a Nd 4$f$ moment,
the differences are larger than $g_{J}J$ of Nd,
which can be seen from the difference between the red open circle and
the red filled circle.
The magnetizations $\mu_{0}M$ are also reduced by $Z$ substitution, as seen from the magnetic moment, but
the amounts of reduction in the magnetizations 
are close to the contribution from the Nd 4$f$ moment.
This is because $Z$ substitution shrinks the volume, and this shrinkage
cancels some of the reduction.
Eventually, the reduction in the magnetization 
falls close to the Nd 4$f$ moment.

For the substitution of Ti for the Fe sites (8$f$, 8$i$, and 8$j$), the magnetic moment is drastically reduced, 
which can be explained by Friedel's concept of
a virtual bound state.~\cite{Fr1958,VeBoZhBu1988,MiTeHaKiIs2014}
As for NdFe$_{11}$Zr and NdFe$_{11}$Hf,
the reduction in $m$ is moderate.
However,
the reduction in terms of 
the magnetization, $\mu_{0}M$, is significantly large
owing to the volume expansion caused by the introduction of Zr and Hf.
%Consequently, $Z$Fe$_{12}$ yields larger magnetization than NdFe$_{11}Z$.

\bibliography{./Reference,./comment_suppl}

%merlin.mbs apsrev4-1.bst 2010-07-25 4.21a (PWD, AO, DPC) hacked
%Control: key (0)
%Control: author (72) initials jnrlst
%Control: editor formatted (1) identically to author
%Control: production of article title (-1) disabled
%Control: page (0) single
%Control: year (1) truncated
%Control: production of eprint (0) enabled
\begin{thebibliography}{39}%
\makeatletter
\providecommand \@ifxundefined [1]{%
 \@ifx{#1\undefined}
}%
\providecommand \@ifnum [1]{%
 \ifnum #1\expandafter \@firstoftwo
 \else \expandafter \@secondoftwo
 \fi
}%
\providecommand \@ifx [1]{%
 \ifx #1\expandafter \@firstoftwo
 \else \expandafter \@secondoftwo
 \fi
}%
\providecommand \natexlab [1]{#1}%
\providecommand \enquote  [1]{``#1''}%
\providecommand \bibnamefont  [1]{#1}%
\providecommand \bibfnamefont [1]{#1}%
\providecommand \citenamefont [1]{#1}%
\providecommand \href@noop [0]{\@secondoftwo}%
\providecommand \href [0]{\begingroup \@sanitize@url \@href}%
\providecommand \@href[1]{\@@startlink{#1}\@@href}%
\providecommand \@@href[1]{\endgroup#1\@@endlink}%
\providecommand \@sanitize@url [0]{\catcode `\\12\catcode `\$12\catcode
  `\&12\catcode `\#12\catcode `\^12\catcode `\_12\catcode `\%12\relax}%
\providecommand \@@startlink[1]{}%
\providecommand \@@endlink[0]{}%
\providecommand \url  [0]{\begingroup\@sanitize@url \@url }%
\providecommand \@url [1]{\endgroup\@href {#1}{\urlprefix }}%
\providecommand \urlprefix  [0]{URL }%
\providecommand \Eprint [0]{\href }%
\providecommand \doibase [0]{http://dx.doi.org/}%
\providecommand \selectlanguage [0]{\@gobble}%
\providecommand \bibinfo  [0]{\@secondoftwo}%
\providecommand \bibfield  [0]{\@secondoftwo}%
\providecommand \translation [1]{[#1]}%
\providecommand \BibitemOpen [0]{}%
\providecommand \bibitemStop [0]{}%
\providecommand \bibitemNoStop [0]{.\EOS\space}%
\providecommand \EOS [0]{\spacefactor3000\relax}%
\providecommand \BibitemShut  [1]{\csname bibitem#1\endcsname}%
\let\auto@bib@innerbib\@empty
%</preamble>
\bibitem [{\citenamefont {Buschow}(1991)}]{Bu1991a}%
  \BibitemOpen
  \bibfield  {author} {\bibinfo {author} {\bibfnamefont {K.~H.~J.}\
  \bibnamefont {Buschow}},\ }\href {\doibase
  http://dx.doi.org/10.1016/0304-8853(91)90813-P} {\bibfield  {journal}
  {\bibinfo  {journal} {J. Magn. Magn. Mater.}\ }\textbf {\bibinfo {volume}
  {100}},\ \bibinfo {pages} {79 } (\bibinfo {year} {1991})}\BibitemShut
  {NoStop}%
\bibitem [{\citenamefont {Hirayama}\ \emph
  {et~al.}(2015{\natexlab{a}})\citenamefont {Hirayama}, \citenamefont
  {Miyake},\ and\ \citenamefont {Hono}}]{HiMiHo2015}%
  \BibitemOpen
  \bibfield  {author} {\bibinfo {author} {\bibfnamefont {Y.}~\bibnamefont
  {Hirayama}}, \bibinfo {author} {\bibfnamefont {T.}~\bibnamefont {Miyake}}, \
  and\ \bibinfo {author} {\bibfnamefont {K.}~\bibnamefont {Hono}},\ }\href
  {\doibase 10.1007/s11837-015-1421-9} {\bibfield  {journal} {\bibinfo
  {journal} {JOM}\ }\textbf {\bibinfo {volume} {67}},\ \bibinfo {pages} {1344}
  (\bibinfo {year} {2015}{\natexlab{a}})}\BibitemShut {NoStop}%
\bibitem [{\citenamefont {Miyake}\ and\ \citenamefont {Akai}(2018)}]{MiAk2018}%
  \BibitemOpen
  \bibfield  {author} {\bibinfo {author} {\bibfnamefont {T.}~\bibnamefont
  {Miyake}}\ and\ \bibinfo {author} {\bibfnamefont {H.}~\bibnamefont {Akai}},\
  }\href {https://doi.org/10.7566/JPSJ.87.041009} {\bibfield  {journal}
  {\bibinfo  {journal} {J. Phys. Soc. Jpn.}\ }\textbf {\bibinfo {volume}
  {87}},\ \bibinfo {pages} {041009} (\bibinfo {year} {2018})}\BibitemShut
  {NoStop}%
\bibitem [{\citenamefont {Miyake}\ \emph {et~al.}(2014)\citenamefont {Miyake},
  \citenamefont {Terakura}, \citenamefont {Harashima}, \citenamefont {Kino},\
  and\ \citenamefont {Ishibashi}}]{MiTeHaKiIs2014}%
  \BibitemOpen
  \bibfield  {author} {\bibinfo {author} {\bibfnamefont {T.}~\bibnamefont
  {Miyake}}, \bibinfo {author} {\bibfnamefont {K.}~\bibnamefont {Terakura}},
  \bibinfo {author} {\bibfnamefont {Y.}~\bibnamefont {Harashima}}, \bibinfo
  {author} {\bibfnamefont {H.}~\bibnamefont {Kino}}, \ and\ \bibinfo {author}
  {\bibfnamefont {S.}~\bibnamefont {Ishibashi}},\ }\href {\doibase
  10.7566/JPSJ.83.043702} {\bibfield  {journal} {\bibinfo  {journal} {J. Phys.
  Soc. Jpn.}\ }\textbf {\bibinfo {volume} {83}},\ \bibinfo {pages} {043702}
  (\bibinfo {year} {2014})}\BibitemShut {NoStop}%
\bibitem [{\citenamefont {Hirayama}\ \emph
  {et~al.}(2015{\natexlab{b}})\citenamefont {Hirayama}, \citenamefont
  {Takahashi}, \citenamefont {Hirosawa},\ and\ \citenamefont
  {Hono}}]{HiTaHiHo2015}%
  \BibitemOpen
  \bibfield  {author} {\bibinfo {author} {\bibfnamefont {Y.}~\bibnamefont
  {Hirayama}}, \bibinfo {author} {\bibfnamefont {Y.~K.}\ \bibnamefont
  {Takahashi}}, \bibinfo {author} {\bibfnamefont {S.}~\bibnamefont {Hirosawa}},
  \ and\ \bibinfo {author} {\bibfnamefont {K.}~\bibnamefont {Hono}},\ }\href
  {\doibase http://dx.doi.org/10.1016/j.scriptamat.2014.10.016} {\bibfield
  {journal} {\bibinfo  {journal} {Scr. Mater.}\ }\textbf {\bibinfo {volume}
  {95}},\ \bibinfo {pages} {70 } (\bibinfo {year}
  {2015}{\natexlab{b}})}\BibitemShut {NoStop}%
\bibitem [{\citenamefont {Yang}\ \emph
  {et~al.}(1991{\natexlab{a}})\citenamefont {Yang}, \citenamefont {Zhang},
  \citenamefont {Kong}, \citenamefont {Pan},\ and\ \citenamefont
  {Ge}}]{YaZhKoPaGe1991}%
  \BibitemOpen
  \bibfield  {author} {\bibinfo {author} {\bibfnamefont {Y.~C.}\ \bibnamefont
  {Yang}}, \bibinfo {author} {\bibfnamefont {X.~D.}\ \bibnamefont {Zhang}},
  \bibinfo {author} {\bibfnamefont {L.~S.}\ \bibnamefont {Kong}}, \bibinfo
  {author} {\bibfnamefont {Q.}~\bibnamefont {Pan}}, \ and\ \bibinfo {author}
  {\bibfnamefont {S.~L.}\ \bibnamefont {Ge}},\ }\href {\doibase
  10.1016/0038-1098(91)90205-A} {\bibfield  {journal} {\bibinfo  {journal}
  {Solid State Commun.}\ }\textbf {\bibinfo {volume} {78}},\ \bibinfo {pages}
  {317 } (\bibinfo {year} {1991}{\natexlab{a}})}\BibitemShut {NoStop}%
\bibitem [{\citenamefont {Yang}\ \emph
  {et~al.}(1991{\natexlab{b}})\citenamefont {Yang}, \citenamefont {Zhang},
  \citenamefont {Ge}, \citenamefont {Pan}, \citenamefont {Kong}, \citenamefont
  {Li}, \citenamefont {Yang}, \citenamefont {Zhang}, \citenamefont {Ding},\
  and\ \citenamefont {Ye}}]{YaZhGePaKoLiYaZhDiYe1991}%
  \BibitemOpen
  \bibfield  {author} {\bibinfo {author} {\bibfnamefont {Y.~C.}\ \bibnamefont
  {Yang}}, \bibinfo {author} {\bibfnamefont {X.~D.}\ \bibnamefont {Zhang}},
  \bibinfo {author} {\bibfnamefont {S.~L.}\ \bibnamefont {Ge}}, \bibinfo
  {author} {\bibfnamefont {Q.}~\bibnamefont {Pan}}, \bibinfo {author}
  {\bibfnamefont {L.~S.}\ \bibnamefont {Kong}}, \bibinfo {author}
  {\bibfnamefont {H.}~\bibnamefont {Li}}, \bibinfo {author} {\bibfnamefont
  {J.~L.}\ \bibnamefont {Yang}}, \bibinfo {author} {\bibfnamefont {B.~S.}\
  \bibnamefont {Zhang}}, \bibinfo {author} {\bibfnamefont {Y.~F.}\ \bibnamefont
  {Ding}}, \ and\ \bibinfo {author} {\bibfnamefont {C.~T.}\ \bibnamefont
  {Ye}},\ }\href {\doibase 10.1063/1.350074} {\bibfield  {journal} {\bibinfo
  {journal} {J. Appl. Phys.}\ }\textbf {\bibinfo {volume} {70}},\ \bibinfo
  {pages} {6001} (\bibinfo {year} {1991}{\natexlab{b}})}\BibitemShut {NoStop}%
\bibitem [{\citenamefont {Harashima}\ \emph {et~al.}(2015)\citenamefont
  {Harashima}, \citenamefont {Terakura}, \citenamefont {Kino}, \citenamefont
  {Ishibashi},\ and\ \citenamefont {Miyake}}]{HaTeKiIsMi2015c}%
  \BibitemOpen
  \bibfield  {author} {\bibinfo {author} {\bibfnamefont {Y.}~\bibnamefont
  {Harashima}}, \bibinfo {author} {\bibfnamefont {K.}~\bibnamefont {Terakura}},
  \bibinfo {author} {\bibfnamefont {H.}~\bibnamefont {Kino}}, \bibinfo {author}
  {\bibfnamefont {S.}~\bibnamefont {Ishibashi}}, \ and\ \bibinfo {author}
  {\bibfnamefont {T.}~\bibnamefont {Miyake}},\ }\href {\doibase
  10.1103/PhysRevB.92.184426} {\bibfield  {journal} {\bibinfo  {journal} {Phys.
  Rev. B}\ }\textbf {\bibinfo {volume} {92}},\ \bibinfo {pages} {184426}
  (\bibinfo {year} {2015})}\BibitemShut {NoStop}%
\bibitem [{\citenamefont {Fukazawa}\ \emph {et~al.}(2017)\citenamefont
  {Fukazawa}, \citenamefont {Akai}, \citenamefont {Harashima},\ and\
  \citenamefont {Miyake}}]{FuAkHaMi2017}%
  \BibitemOpen
  \bibfield  {author} {\bibinfo {author} {\bibfnamefont {T.}~\bibnamefont
  {Fukazawa}}, \bibinfo {author} {\bibfnamefont {H.}~\bibnamefont {Akai}},
  \bibinfo {author} {\bibfnamefont {Y.}~\bibnamefont {Harashima}}, \ and\
  \bibinfo {author} {\bibfnamefont {T.}~\bibnamefont {Miyake}},\ }\href
  {\doibase 10.1063/1.4996989} {\bibfield  {journal} {\bibinfo  {journal} {J.
  Appl. Phys.}\ }\textbf {\bibinfo {volume} {122}},\ \bibinfo {pages} {053901}
  (\bibinfo {year} {2017})}\BibitemShut {NoStop}%
\bibitem [{\citenamefont {Coehoorn}(1990)}]{Co1990}%
  \BibitemOpen
  \bibfield  {author} {\bibinfo {author} {\bibfnamefont {R.}~\bibnamefont
  {Coehoorn}},\ }\href {\doibase 10.1103/PhysRevB.41.11790} {\bibfield
  {journal} {\bibinfo  {journal} {Phys. Rev. B}\ }\textbf {\bibinfo {volume}
  {41}},\ \bibinfo {pages} {11790} (\bibinfo {year} {1990})}\BibitemShut
  {NoStop}%
\bibitem [{\citenamefont {Margarian}\ \emph {et~al.}(1994)\citenamefont
  {Margarian}, \citenamefont {Dunlop}, \citenamefont {Day},\ and\ \citenamefont
  {Kalceff}}]{MaDuDaKa1994}%
  \BibitemOpen
  \bibfield  {author} {\bibinfo {author} {\bibfnamefont {A.}~\bibnamefont
  {Margarian}}, \bibinfo {author} {\bibfnamefont {J.~B.}\ \bibnamefont
  {Dunlop}}, \bibinfo {author} {\bibfnamefont {R.~K.}\ \bibnamefont {Day}}, \
  and\ \bibinfo {author} {\bibfnamefont {W.}~\bibnamefont {Kalceff}},\ }\href
  {\doibase http://dx.doi.org/10.1063/1.358338} {\bibfield  {journal} {\bibinfo
   {journal} {J. Appl. Phys.}\ }\textbf {\bibinfo {volume} {76}},\ \bibinfo
  {pages} {6153} (\bibinfo {year} {1994})}\BibitemShut {NoStop}%
\bibitem [{\citenamefont {Suzuki}(2017)}]{Su2017}%
  \BibitemOpen
  \bibfield  {author} {\bibinfo {author} {\bibfnamefont {H.}~\bibnamefont
  {Suzuki}},\ }\href {https://doi.org/10.1063/1.4973799} {\bibfield  {journal}
  {\bibinfo  {journal} {AIP Adv.}\ }\textbf {\bibinfo {volume} {7}},\ \bibinfo
  {pages} {056208} (\bibinfo {year} {2017})}\BibitemShut {NoStop}%
\bibitem [{\citenamefont {Ohashi}\ \emph
  {et~al.}(1988{\natexlab{a}})\citenamefont {Ohashi}, \citenamefont {Tawara},
  \citenamefont {Osugi}, \citenamefont {Sakurai},\ and\ \citenamefont
  {Komura}}]{OhTaOsSaKo1988}%
  \BibitemOpen
  \bibfield  {author} {\bibinfo {author} {\bibfnamefont {K.}~\bibnamefont
  {Ohashi}}, \bibinfo {author} {\bibfnamefont {Y.}~\bibnamefont {Tawara}},
  \bibinfo {author} {\bibfnamefont {R.}~\bibnamefont {Osugi}}, \bibinfo
  {author} {\bibfnamefont {J.}~\bibnamefont {Sakurai}}, \ and\ \bibinfo
  {author} {\bibfnamefont {Y.}~\bibnamefont {Komura}},\ }\href {\doibase
  10.1016/0022-5088(88)90020-3} {\bibfield  {journal} {\bibinfo  {journal} {J.
  Less-Common Met.}\ }\textbf {\bibinfo {volume} {139}},\ \bibinfo {pages} {L1
  } (\bibinfo {year} {1988}{\natexlab{a}})}\BibitemShut {NoStop}%
\bibitem [{\citenamefont {Ohashi}\ \emph
  {et~al.}(1988{\natexlab{b}})\citenamefont {Ohashi}, \citenamefont {Tawara},
  \citenamefont {Osugi},\ and\ \citenamefont {Shimao}}]{OhTaOsSh1988}%
  \BibitemOpen
  \bibfield  {author} {\bibinfo {author} {\bibfnamefont {K.}~\bibnamefont
  {Ohashi}}, \bibinfo {author} {\bibfnamefont {Y.}~\bibnamefont {Tawara}},
  \bibinfo {author} {\bibfnamefont {R.}~\bibnamefont {Osugi}}, \ and\ \bibinfo
  {author} {\bibfnamefont {M.}~\bibnamefont {Shimao}},\ }\href {\doibase
  http://dx.doi.org/10.1063/1.342235} {\bibfield  {journal} {\bibinfo
  {journal} {J. Appl. Phys.}\ }\textbf {\bibinfo {volume} {64}},\ \bibinfo
  {pages} {5714} (\bibinfo {year} {1988}{\natexlab{b}})}\BibitemShut {NoStop}%
\bibitem [{\citenamefont {Hirosawa}\ \emph {et~al.}(2017)\citenamefont
  {Hirosawa}, \citenamefont {Nishino},\ and\ \citenamefont
  {Miyashita}}]{HiNiMi2017}%
  \BibitemOpen
  \bibfield  {author} {\bibinfo {author} {\bibfnamefont {S.}~\bibnamefont
  {Hirosawa}}, \bibinfo {author} {\bibfnamefont {M.}~\bibnamefont {Nishino}}, \
  and\ \bibinfo {author} {\bibfnamefont {S.}~\bibnamefont {Miyashita}},\ }\href
  {http://stacks.iop.org/2043-6262/8/i=1/a=013002} {\bibfield  {journal}
  {\bibinfo  {journal} {Adv. Nat. Sci.: Nanosci. Nanotechnol.}\ }\textbf
  {\bibinfo {volume} {8}},\ \bibinfo {pages} {013002} (\bibinfo {year}
  {2017})}\BibitemShut {NoStop}%
\bibitem [{\citenamefont {Felner}(1980)}]{Fe1980}%
  \BibitemOpen
  \bibfield  {author} {\bibinfo {author} {\bibfnamefont {I.}~\bibnamefont
  {Felner}},\ }\href {\doibase http://dx.doi.org/10.1016/0022-5088(80)90143-5}
  {\bibfield  {journal} {\bibinfo  {journal} {J. Less-Common Met.}\ }\textbf
  {\bibinfo {volume} {72}},\ \bibinfo {pages} {241 } (\bibinfo {year}
  {1980})}\BibitemShut {NoStop}%
\bibitem [{\citenamefont {Yang}\ \emph {et~al.}(1981)\citenamefont {Yang},
  \citenamefont {Kebe}, \citenamefont {James}, \citenamefont {Deportes},\ and\
  \citenamefont {Yelon}}]{YaKeJaDeYe1981}%
  \BibitemOpen
  \bibfield  {author} {\bibinfo {author} {\bibfnamefont {Y.~C.}\ \bibnamefont
  {Yang}}, \bibinfo {author} {\bibfnamefont {B.}~\bibnamefont {Kebe}}, \bibinfo
  {author} {\bibfnamefont {W.~J.}\ \bibnamefont {James}}, \bibinfo {author}
  {\bibfnamefont {J.}~\bibnamefont {Deportes}}, \ and\ \bibinfo {author}
  {\bibfnamefont {W.}~\bibnamefont {Yelon}},\ }\href {\doibase
  http://dx.doi.org/10.1063/1.329621} {\bibfield  {journal} {\bibinfo
  {journal} {J. Appl. Phys.}\ }\textbf {\bibinfo {volume} {52}},\ \bibinfo
  {pages} {2077} (\bibinfo {year} {1981})}\BibitemShut {NoStop}%
\bibitem [{\citenamefont {Mooij}\ and\ \citenamefont
  {Buschow}(1988)}]{MoBu1988}%
  \BibitemOpen
  \bibfield  {author} {\bibinfo {author} {\bibfnamefont {D.~B.~D.}\
  \bibnamefont {Mooij}}\ and\ \bibinfo {author} {\bibfnamefont {K.~H.~J.}\
  \bibnamefont {Buschow}},\ }\href {\doibase
  http://dx.doi.org/10.1016/0022-5088(88)90424-9} {\bibfield  {journal}
  {\bibinfo  {journal} {J. Less-Common Met.}\ }\textbf {\bibinfo {volume}
  {136}},\ \bibinfo {pages} {207 } (\bibinfo {year} {1988})}\BibitemShut
  {NoStop}%
\bibitem [{\citenamefont {M{\"u}ller}(1988)}]{Mu1988}%
  \BibitemOpen
  \bibfield  {author} {\bibinfo {author} {\bibfnamefont {A.}~\bibnamefont
  {M{\"u}ller}},\ }\href {\doibase http://dx.doi.org/10.1063/1.341473}
  {\bibfield  {journal} {\bibinfo  {journal} {J. Appl. Phys.}\ }\textbf
  {\bibinfo {volume} {64}},\ \bibinfo {pages} {249} (\bibinfo {year}
  {1988})}\BibitemShut {NoStop}%
\bibitem [{\citenamefont {Wang}\ \emph {et~al.}(1988)\citenamefont {Wang},
  \citenamefont {Chevalier}, \citenamefont {Berlureau}, \citenamefont
  {Etourneau}, \citenamefont {Coey},\ and\ \citenamefont
  {Cadogan}}]{WaChBeEtCoCa1988}%
  \BibitemOpen
  \bibfield  {author} {\bibinfo {author} {\bibfnamefont {X.~Z.}\ \bibnamefont
  {Wang}}, \bibinfo {author} {\bibfnamefont {B.}~\bibnamefont {Chevalier}},
  \bibinfo {author} {\bibfnamefont {T.}~\bibnamefont {Berlureau}}, \bibinfo
  {author} {\bibfnamefont {J.}~\bibnamefont {Etourneau}}, \bibinfo {author}
  {\bibfnamefont {J.~M.~D.}\ \bibnamefont {Coey}}, \ and\ \bibinfo {author}
  {\bibfnamefont {J.~M.}\ \bibnamefont {Cadogan}},\ }\href {\doibase
  http://dx.doi.org/10.1016/0022-5088(88)90112-9} {\bibfield  {journal}
  {\bibinfo  {journal} {J. Less-Common Met.}\ }\textbf {\bibinfo {volume}
  {138}},\ \bibinfo {pages} {235 } (\bibinfo {year} {1988})}\BibitemShut
  {NoStop}%
\bibitem [{\citenamefont {Harashima}\ \emph {et~al.}(2016)\citenamefont
  {Harashima}, \citenamefont {Terakura}, \citenamefont {Kino}, \citenamefont
  {Ishibashi},\ and\ \citenamefont {Miyake}}]{HaTeKiIsMi2016}%
  \BibitemOpen
  \bibfield  {author} {\bibinfo {author} {\bibfnamefont {Y.}~\bibnamefont
  {Harashima}}, \bibinfo {author} {\bibfnamefont {K.}~\bibnamefont {Terakura}},
  \bibinfo {author} {\bibfnamefont {H.}~\bibnamefont {Kino}}, \bibinfo {author}
  {\bibfnamefont {S.}~\bibnamefont {Ishibashi}}, \ and\ \bibinfo {author}
  {\bibfnamefont {T.}~\bibnamefont {Miyake}},\ }\href
  {http://scitation.aip.org/content/aip/journal/jap/120/20/10.1063/1.4968798}
  {\bibfield  {journal} {\bibinfo  {journal} {J. Appl. Phys.}\ }\textbf
  {\bibinfo {volume} {120}},\ \bibinfo {eid} {203904} (\bibinfo {year}
  {2016})}\BibitemShut {NoStop}%
\bibitem [{\citenamefont {Hirayama}\ \emph {et~al.}(2017)\citenamefont
  {Hirayama}, \citenamefont {Takahashi}, \citenamefont {Hirosawa},\ and\
  \citenamefont {Hono}}]{HiTaHiHo2017}%
  \BibitemOpen
  \bibfield  {author} {\bibinfo {author} {\bibfnamefont {Y.}~\bibnamefont
  {Hirayama}}, \bibinfo {author} {\bibfnamefont {Y.~K.}\ \bibnamefont
  {Takahashi}}, \bibinfo {author} {\bibfnamefont {S.}~\bibnamefont {Hirosawa}},
  \ and\ \bibinfo {author} {\bibfnamefont {K.}~\bibnamefont {Hono}},\ }\href
  {\doibase https://doi.org/10.1016/j.scriptamat.2017.05.029} {\bibfield
  {journal} {\bibinfo  {journal} {Scr. Mater.}\ }\textbf {\bibinfo {volume}
  {138}},\ \bibinfo {pages} {62 } (\bibinfo {year} {2017})}\BibitemShut
  {NoStop}%
\bibitem [{\citenamefont {Suzuki}\ \emph {et~al.}(2014)\citenamefont {Suzuki},
  \citenamefont {Kuno}, \citenamefont {Urushibata}, \citenamefont {Kobayashi},
  \citenamefont {Sakuma}, \citenamefont {Washio}, \citenamefont {Kishimoto},
  \citenamefont {Kato},\ and\ \citenamefont {Manabe}}]{SuKuUrKoSaWaKiKaMa2014}%
  \BibitemOpen
  \bibfield  {author} {\bibinfo {author} {\bibfnamefont {S.}~\bibnamefont
  {Suzuki}}, \bibinfo {author} {\bibfnamefont {T.}~\bibnamefont {Kuno}},
  \bibinfo {author} {\bibfnamefont {K.}~\bibnamefont {Urushibata}}, \bibinfo
  {author} {\bibfnamefont {K.}~\bibnamefont {Kobayashi}}, \bibinfo {author}
  {\bibfnamefont {N.}~\bibnamefont {Sakuma}}, \bibinfo {author} {\bibfnamefont
  {K.}~\bibnamefont {Washio}}, \bibinfo {author} {\bibfnamefont
  {H.}~\bibnamefont {Kishimoto}}, \bibinfo {author} {\bibfnamefont
  {A.}~\bibnamefont {Kato}}, \ and\ \bibinfo {author} {\bibfnamefont
  {A.}~\bibnamefont {Manabe}},\ }\href
  {http://scitation.aip.org/content/aip/journal/adva/4/11/10.1063/1.4902176}
  {\bibfield  {journal} {\bibinfo  {journal} {AIP Adv.}\ }\textbf {\bibinfo
  {volume} {4}},\ \bibinfo {eid} {117131} (\bibinfo {year} {2014})}\BibitemShut
  {NoStop}%
\bibitem [{\citenamefont {Sakuma}\ \emph {et~al.}(2016)\citenamefont {Sakuma},
  \citenamefont {Suzuki}, \citenamefont {Kuno}, \citenamefont {Urushibata},
  \citenamefont {Kobayashi}, \citenamefont {Yano}, \citenamefont {Kato},\ and\
  \citenamefont {Manabe}}]{SaSuKuUrKoYaKaMa2016}%
  \BibitemOpen
  \bibfield  {author} {\bibinfo {author} {\bibfnamefont {N.}~\bibnamefont
  {Sakuma}}, \bibinfo {author} {\bibfnamefont {S.}~\bibnamefont {Suzuki}},
  \bibinfo {author} {\bibfnamefont {T.}~\bibnamefont {Kuno}}, \bibinfo {author}
  {\bibfnamefont {K.}~\bibnamefont {Urushibata}}, \bibinfo {author}
  {\bibfnamefont {K.}~\bibnamefont {Kobayashi}}, \bibinfo {author}
  {\bibfnamefont {M.}~\bibnamefont {Yano}}, \bibinfo {author} {\bibfnamefont
  {A.}~\bibnamefont {Kato}}, \ and\ \bibinfo {author} {\bibfnamefont
  {A.}~\bibnamefont {Manabe}},\ }\href {http://dx.doi.org/10.1063/1.4944521}
  {\bibfield  {journal} {\bibinfo  {journal} {AIP Adv.}\ }\textbf {\bibinfo
  {volume} {6}},\ \bibinfo {pages} {056023} (\bibinfo {year}
  {2016})}\BibitemShut {NoStop}%
\bibitem [{\citenamefont {Kuno}\ \emph {et~al.}(2016)\citenamefont {Kuno},
  \citenamefont {Suzuki}, \citenamefont {Urushibata}, \citenamefont
  {Kobayashi}, \citenamefont {Sakuma}, \citenamefont {Yano}, \citenamefont
  {Kato},\ and\ \citenamefont {Manabe}}]{KuSuUrKoSaYaKaMa2016}%
  \BibitemOpen
  \bibfield  {author} {\bibinfo {author} {\bibfnamefont {T.}~\bibnamefont
  {Kuno}}, \bibinfo {author} {\bibfnamefont {S.}~\bibnamefont {Suzuki}},
  \bibinfo {author} {\bibfnamefont {K.}~\bibnamefont {Urushibata}}, \bibinfo
  {author} {\bibfnamefont {K.}~\bibnamefont {Kobayashi}}, \bibinfo {author}
  {\bibfnamefont {N.}~\bibnamefont {Sakuma}}, \bibinfo {author} {\bibfnamefont
  {M.}~\bibnamefont {Yano}}, \bibinfo {author} {\bibfnamefont {A.}~\bibnamefont
  {Kato}}, \ and\ \bibinfo {author} {\bibfnamefont {A.}~\bibnamefont
  {Manabe}},\ }\href
  {http://scitation.aip.org/content/aip/journal/adva/6/2/10.1063/1.4943051}
  {\bibfield  {journal} {\bibinfo  {journal} {AIP Adv.}\ }\textbf {\bibinfo
  {volume} {6}},\ \bibinfo {eid} {025221} (\bibinfo {year} {2016})}\BibitemShut
  {NoStop}%
\bibitem [{\citenamefont {Suzuki}\ \emph {et~al.}(2016)\citenamefont {Suzuki},
  \citenamefont {Kuno}, \citenamefont {Urushibata}, \citenamefont {Kobayashi},
  \citenamefont {Sakuma}, \citenamefont {Washio}, \citenamefont {Yano},
  \citenamefont {Kato},\ and\ \citenamefont {Manabe}}]{SuKuUrKoSaWaYaKaMa2016}%
  \BibitemOpen
  \bibfield  {author} {\bibinfo {author} {\bibfnamefont {S.}~\bibnamefont
  {Suzuki}}, \bibinfo {author} {\bibfnamefont {T.}~\bibnamefont {Kuno}},
  \bibinfo {author} {\bibfnamefont {K.}~\bibnamefont {Urushibata}}, \bibinfo
  {author} {\bibfnamefont {K.}~\bibnamefont {Kobayashi}}, \bibinfo {author}
  {\bibfnamefont {N.}~\bibnamefont {Sakuma}}, \bibinfo {author} {\bibfnamefont
  {K.}~\bibnamefont {Washio}}, \bibinfo {author} {\bibfnamefont
  {M.}~\bibnamefont {Yano}}, \bibinfo {author} {\bibfnamefont {A.}~\bibnamefont
  {Kato}}, \ and\ \bibinfo {author} {\bibfnamefont {A.}~\bibnamefont
  {Manabe}},\ }\href {\doibase http://dx.doi.org/10.1016/j.jmmm.2015.10.042}
  {\bibfield  {journal} {\bibinfo  {journal} {J. Magn. Magn. Mater.}\ }\textbf
  {\bibinfo {volume} {401}},\ \bibinfo {pages} {259 } (\bibinfo {year}
  {2016})}\BibitemShut {NoStop}%
\bibitem [{Qm2()}]{Qm2014}%
  \BibitemOpen
  \href@noop {} {}\bibinfo {howpublished} {\url{http://qmas.jp/}}\BibitemShut
  {NoStop}%
\bibitem [{\citenamefont {Hohenberg}\ and\ \citenamefont
  {Kohn}(1964)}]{HoKo1964}%
  \BibitemOpen
  \bibfield  {author} {\bibinfo {author} {\bibfnamefont {P.}~\bibnamefont
  {Hohenberg}}\ and\ \bibinfo {author} {\bibfnamefont {W.}~\bibnamefont
  {Kohn}},\ }\href {\doibase 10.1103/PhysRev.136.B864} {\bibfield  {journal}
  {\bibinfo  {journal} {Phys. Rev.}\ }\textbf {\bibinfo {volume} {136}},\
  \bibinfo {pages} {B864} (\bibinfo {year} {1964})}\BibitemShut {NoStop}%
\bibitem [{\citenamefont {Kohn}\ and\ \citenamefont {Sham}(1965)}]{KoSh1965}%
  \BibitemOpen
  \bibfield  {author} {\bibinfo {author} {\bibfnamefont {W.}~\bibnamefont
  {Kohn}}\ and\ \bibinfo {author} {\bibfnamefont {L.~J.}\ \bibnamefont
  {Sham}},\ }\href {\doibase 10.1103/PhysRev.140.A1133} {\bibfield  {journal}
  {\bibinfo  {journal} {Phys. Rev.}\ }\textbf {\bibinfo {volume} {140}},\
  \bibinfo {pages} {A1133} (\bibinfo {year} {1965})}\BibitemShut {NoStop}%
\bibitem [{\citenamefont {Bl\"ochl}(1994)}]{Bl1994}%
  \BibitemOpen
  \bibfield  {author} {\bibinfo {author} {\bibfnamefont {P.~E.}\ \bibnamefont
  {Bl\"ochl}},\ }\href {\doibase 10.1103/PhysRevB.50.17953} {\bibfield
  {journal} {\bibinfo  {journal} {Phys. Rev. B}\ }\textbf {\bibinfo {volume}
  {50}},\ \bibinfo {pages} {17953} (\bibinfo {year} {1994})}\BibitemShut
  {NoStop}%
\bibitem [{\citenamefont {Kresse}\ and\ \citenamefont
  {Joubert}(1999)}]{KrJo1999}%
  \BibitemOpen
  \bibfield  {author} {\bibinfo {author} {\bibfnamefont {G.}~\bibnamefont
  {Kresse}}\ and\ \bibinfo {author} {\bibfnamefont {D.}~\bibnamefont
  {Joubert}},\ }\href {\doibase 10.1103/PhysRevB.59.1758} {\bibfield  {journal}
  {\bibinfo  {journal} {Phys. Rev. B}\ }\textbf {\bibinfo {volume} {59}},\
  \bibinfo {pages} {1758} (\bibinfo {year} {1999})}\BibitemShut {NoStop}%
\bibitem [{\citenamefont {Perdew}\ \emph {et~al.}(1996)\citenamefont {Perdew},
  \citenamefont {Burke},\ and\ \citenamefont {Ernzerhof}}]{PeBuEr1996}%
  \BibitemOpen
  \bibfield  {author} {\bibinfo {author} {\bibfnamefont {J.~P.}\ \bibnamefont
  {Perdew}}, \bibinfo {author} {\bibfnamefont {K.}~\bibnamefont {Burke}}, \
  and\ \bibinfo {author} {\bibfnamefont {M.}~\bibnamefont {Ernzerhof}},\ }\href
  {\doibase 10.1103/PhysRevLett.77.3865} {\bibfield  {journal} {\bibinfo
  {journal} {Phys. Rev. Lett.}\ }\textbf {\bibinfo {volume} {77}},\ \bibinfo
  {pages} {3865} (\bibinfo {year} {1996})}\BibitemShut {NoStop}%
\bibitem [{\citenamefont {Tatetsu}\ \emph {et~al.}(2018)\citenamefont
  {Tatetsu}, \citenamefont {Harashima}, \citenamefont {Miyake},\ and\
  \citenamefont {Gohda}}]{TaHaMiGo2018}%
  \BibitemOpen
  \bibfield  {author} {\bibinfo {author} {\bibfnamefont {Y.}~\bibnamefont
  {Tatetsu}}, \bibinfo {author} {\bibfnamefont {Y.}~\bibnamefont {Harashima}},
  \bibinfo {author} {\bibfnamefont {T.}~\bibnamefont {Miyake}}, \ and\ \bibinfo
  {author} {\bibfnamefont {Y.}~\bibnamefont {Gohda}},\ }\href {\doibase
  10.1103/PhysRevMaterials.2.074410} {\bibfield  {journal} {\bibinfo  {journal}
  {Phys. Rev. Materials}\ }\textbf {\bibinfo {volume} {2}},\ \bibinfo {pages}
  {074410} (\bibinfo {year} {2018})}\BibitemShut {NoStop}%
\bibitem [{\citenamefont {Coey}\ and\ \citenamefont {Sun}(1990)}]{CoSu1990}%
  \BibitemOpen
  \bibfield  {author} {\bibinfo {author} {\bibfnamefont {J.~M.~D.}\
  \bibnamefont {Coey}}\ and\ \bibinfo {author} {\bibfnamefont {H.}~\bibnamefont
  {Sun}},\ }\href {\doibase http://dx.doi.org/10.1016/0304-8853(90)90756-G}
  {\bibfield  {journal} {\bibinfo  {journal} {J. Magn. Magn. Mater.}\ }\textbf
  {\bibinfo {volume} {87}},\ \bibinfo {pages} {L251 } (\bibinfo {year}
  {1990})}\BibitemShut {NoStop}%
\bibitem [{\citenamefont {Koyama}\ and\ \citenamefont
  {Fujii}(2000)}]{KoFu2000}%
  \BibitemOpen
  \bibfield  {author} {\bibinfo {author} {\bibfnamefont {K.}~\bibnamefont
  {Koyama}}\ and\ \bibinfo {author} {\bibfnamefont {H.}~\bibnamefont {Fujii}},\
  }\href {\doibase 10.1103/PhysRevB.61.9475} {\bibfield  {journal} {\bibinfo
  {journal} {Phys. Rev. B}\ }\textbf {\bibinfo {volume} {61}},\ \bibinfo
  {pages} {9475} (\bibinfo {year} {2000})}\BibitemShut {NoStop}%
\bibitem [{\citenamefont {Hirayama}\ \emph {et~al.}(2016)\citenamefont
  {Hirayama}, \citenamefont {Panda}, \citenamefont {Ohkubo},\ and\
  \citenamefont {Hono}}]{HiPaOhHo2016}%
  \BibitemOpen
  \bibfield  {author} {\bibinfo {author} {\bibfnamefont {Y.}~\bibnamefont
  {Hirayama}}, \bibinfo {author} {\bibfnamefont {A.~K.}\ \bibnamefont {Panda}},
  \bibinfo {author} {\bibfnamefont {T.}~\bibnamefont {Ohkubo}}, \ and\ \bibinfo
  {author} {\bibfnamefont {K.}~\bibnamefont {Hono}},\ }\href {\doibase
  https://doi.org/10.1016/j.scriptamat.2016.03.028} {\bibfield  {journal}
  {\bibinfo  {journal} {Scr. Mater.}\ }\textbf {\bibinfo {volume} {120}},\
  \bibinfo {pages} {27 } (\bibinfo {year} {2016})}\BibitemShut {NoStop}%
\bibitem [{sup()}]{supplementalmaterial}%
  \BibitemOpen
  \href@noop {} {}\bibinfo {note} {See Supplemental Material for (SA) the
  lattice constants of $R$Fe$_{12}$ and $R_{2}$Fe$_{17}$, (SB) the calculated
  atomic radii of $R$ and Fe, (SC) the difference in the total energies of
  rhombohedral and hexagonal $R_{2}$Fe$_{17}$, (SD) the bond lengths in
  $R$Fe$_{12}$, and (SE) the lattice distortion in NdFe$_{12}$ and
  Nd$_{2}$Fe$_{17}$ induced by hydrostatic pressure.}\BibitemShut {Stop}%
\bibitem [{\citenamefont {Friedel}(1958)}]{Fr1958}%
  \BibitemOpen
  \bibfield  {author} {\bibinfo {author} {\bibfnamefont {J.}~\bibnamefont
  {Friedel}},\ }\href {\doibase 10.1007/BF02751483} {\bibfield  {journal}
  {\bibinfo  {journal} {Il Nuovo Cimento (1955-1965)}\ }\textbf {\bibinfo
  {volume} {7}},\ \bibinfo {pages} {287} (\bibinfo {year} {1958})}\BibitemShut
  {NoStop}%
\bibitem [{\citenamefont {Verhoef}\ \emph {et~al.}(1988)\citenamefont
  {Verhoef}, \citenamefont {de~Boer}, \citenamefont {Zhi-dong},\ and\
  \citenamefont {Buschow}}]{VeBoZhBu1988}%
  \BibitemOpen
  \bibfield  {author} {\bibinfo {author} {\bibfnamefont {R.}~\bibnamefont
  {Verhoef}}, \bibinfo {author} {\bibfnamefont {F.~R.}\ \bibnamefont
  {de~Boer}}, \bibinfo {author} {\bibfnamefont {Z.}~\bibnamefont {Zhi-dong}}, \
  and\ \bibinfo {author} {\bibfnamefont {K.~H.~J.}\ \bibnamefont {Buschow}},\
  }\href {\doibase http://dx.doi.org/10.1016/0304-8853(88)90037-6} {\bibfield
  {journal} {\bibinfo  {journal} {J. Magn. Magn. Mater.}\ }\textbf {\bibinfo
  {volume} {75}},\ \bibinfo {pages} {319 } (\bibinfo {year}
  {1988})}\BibitemShut {NoStop}%
\end{thebibliography}%

\end{document}